\documentstyle[11pt]{article}

\newcommand{\TabCap}[2]{\begin{center}\parbox[t]{#1}{\begin{center}
  \small {\spaceskip 2pt plus 1pt minus 1pt T a b l e}
  \refstepcounter{table}\thetable \\[2mm]
  \footnotesize #2 \end{center}}\end{center}}
\newcommand{\TableFont}{\footnotesize}
\newcommand{\MakeTable}[4]{\begin{table}[htb]\TabCap{#2}{#3}
  \begin{center} \TableFont \begin{tabular}{#1} #4 
  \end{tabular}\end{center}\end{table}}
\newcommand{\vs}{{\it vs.\ }}
\newcommand{\etal}{{\it et al.\ }}
\newcommand{\trule}{\rule{0pt}{14pt}}
\newcommand{\zdot}{\makebox[0pt][l]{.}}
\newcommand{\up}[1]{\ifmmode^{\rm #1}\else$^{\rm #1}$\fi}

\newcommand{\uph}{\up{h}}
\newcommand{\upm}{\up{m}}
\newcommand{\ups}{\up{s}}
\newcommand{\arcd}{\ifmmode^{\circ}\else$^{\circ}$\fi}
\newcommand{\arcm}{\ifmmode{'}\else$'$\fi}
\newcommand{\arcs}{\ifmmode{''}\else$''$\fi}

\newenvironment{references}%
{
\footnotesize \frenchspacing

\newcommand{\ApJ}{Astrophys.\ J.}
\newcommand{\ApJS}{Astrophys.\ J.~Suppl.~Ser.}
\newcommand{\ApJL}{Astrophys.\ J.~Letters}
\newcommand{\AJ}{Astron.\ J.}

\newcommand{\PASP}{P.A.S.P.}
\newcommand{\Acta}{Acta Astron.}

\renewcommand{\and}{{\rm and }}
\section{{\rm REFERENCES}}
\sloppy \hyphenpenalty10000
\begin{list}{}{\leftmargin1cm\listparindent-1cm
\itemindent\listparindent\parsep0pt\itemsep0pt}}%
{\end{list}\vspace{2mm}}

\def\TYLDA{~}
\newlength{\DW}
\newcommand{\refitem}[5]{\item[]{#1} #2%
\def\REFARG{#3}\ifx\REFARG\TYLDA\else, {\it#3}\fi
\def\REFARG{#4}\ifx\REFARG\TYLDA\else, {\bf#4}\fi
\def\REFARG{#5}\ifx\REFARG\TYLDA\else, {#5}\fi.}

\begin{document}

\def\thefootnote{\fnsymbol{footnote}}

\begin{center}
\large\bf
The Optical Gravitational Lensing Experiment.\\ 
The General Catalog of Stars in the Galactic Bulge.\\
I. The Stars in the Central Baade's Window OGLE Field BWC\footnote{Based on 
observations obtained at the Las Campanas Observatory of the Carnegie 
Institution of Washington} \\[5mm]
{\it by} \\[5mm]
\def\thefootnote{\arabic{footnote}}
\large
~M.~~S~z~y~m~a~\'n~s~k~i$^1$,
A.~~U~d~a~l~s~k~i$^1$,
~M.~~K~u~b~i~a~k$^1$,
~J.~~K~a~\l~u~\.z~n~y$^1$,
~M.~~M~a~t~e~o$^2$
~~and~~W.~~K~r~z~e~m~i~\'n~s~k~i$^3$ \\[2mm]
\normalsize
$^1$Warsaw University Observatory, 
Al.~Ujazdowskie~4, 00--478~Warszawa, Poland\\
e-mail: (msz,udalski,mk,jka)@sirius.astrouw.edu.pl\\
\vskip9pt
$^2$Department of Astronomy, University of Michigan, 821~Dennison 
Bldg., Ann Arbor, MI~48109--1090, USA\\
e-mail: mateo@astro.lsa.umich.edu\\
\vskip3pt
$^3$Carnegie Observatories, Las Campanas Observatory, Casilla~601, 
La~Serena, Chile\\
e-mail: wojtek@roses.ctio.noao.edu
\end{center}

{\bf Abstract.}
This paper presents the first part of the Optical Gravitational 
Lensing Experiment (OGLE) General Catalog of Stars
in the Galactic bulge. The Catalog is based on observations collected
during the OGLE microlensing search. This part contains 33196 stars
brighter than ${I=18\upm}$ identified in the Baade's Window BWC field.
Stars from remaining 20 OGLE fields will be presented in similar form in
the next parts of the Catalog. The Catalog is available to the
astronomical community over the Internet network.

\section{Introduction}
The Optical Gravitational Lensing Experiment (OGLE) is a long term
observing  project with the main goal of searching for dark  matter in
our Galaxy using microlensing (Paczy\'nski 1986). The fourth season of
observations, 1995, concluded the first phase of the OGLE project. It
will be resumed in 1996 using a dedicated 1.3~m Warsaw photometric
telescope installed at Las Campanas Observatory in February 1996.

Four seasons of continuous photometric monitoring of approximately
two million non-variable stars in the direction of the Galactic bulge
resulted in discovery of 18 possible microlensing events (Udalski \etal
1993b, 1994b, 1994c). 
\vskip4pt
The collected photometric data provide a unique material for
studying the Galactic bulge (Paczy\'nski \etal 1994, Stanek \etal 1994).
Long span and large number of observations make it also possible to
study precisely the population of variable stars toward the Galactic
bulge. Three parts of the OGLE Periodic Variable Stars Catalog have
already been published (Udalski \etal 1994d, 1995a, 1995c).
\vskip4pt
This paper is the opening of a series of papers constituting the OGLE
General Catalog of Stars in the Galactic bulge. The aim of the Catalog
is to provide possibly most complete reference to objects in the areas
of the Galactic bulge observed regularly during the OGLE microlensing
search.  The Catalog is meant to be an open publication with periodic
updates when more data become available and when
fainter stars are included. We hope this catalog will be
useful for all kind of studies of the objects in the Galactic
bulge.
\vskip4pt
In this paper we present 33196 stars detected in one of the OGLE
fields designated as BWC, that is in the center of the Baade's Window. In
the following papers the stars from remaining 11 Baade's Window fields will
be presented, followed by data from 9 fields in other parts of the
Galactic bulge. 
\vskip4pt
In Section~2 we describe shortly the reduction technique. Section~3
presents the Catalog and gives a sample of its data. The whole set of
data, much too big to be published in a journal, is available in the
electronic form (see Section~5 for details). Section~4 gives a short
discussion of the completeness of the Catalog.

\vspace*{20pt}
\section{Reductions}
The OGLE General Catalog of Stars in the Galactic bulge is based on
observations obtained during the OGLE microlensing search. Data
presented here were collected during four OGLE observing seasons, 1992
-- 1995, spanning the period from April 19, 1992  to August 22, 1995.
Each observing season lasted approximately from April to
August/mid-September. Full logs of observations can be found in Udalski
\etal (1992), Udalski \etal (1994a), Udalski \etal (1995b) and Udalski
\etal (1996).
\vskip4pt
All observations were collected with the 1-m Swope telescope at Las
Campanas Observatory which is operated by the Carnegie Institution of
Washington. ${2048\times2048}$ Ford/Loral CCD with 15~$\mu$m pixels
giving a scale of 0.44~arcsec/pixel was used as the detector. Due to
the strategy adopted for microlensing search the majority of
observations was made in the $I$-band with much less $V$-band
measurements taken to obtain color estimates.

All the collected frames were reduced using the standard OGLE
data-pipeline in which, after automatic debiasing and flat-fielding, the
profile photometry of objects was derived. Reductions of frames were
done in the near real time using the modified DoPHOT photometry program
in the fixed position mode (Schechter \etal 1993). Due to significant
variations of the Point Spread Function (PSF) over the chip, each frame
was divided into a ${7\times7}$ grid of slightly overlapping subframes on
which PSF function could be assumed as constant. The photometry from all
subframes was then tied to the common photometric system using stars
measured on the overlapping parts of the subframes. Depending on the
quality of the photometry a grade {\bf A} -- {\bf F} was assigned to
each frame with {\bf A} meaning the best and {\bf F} -- unacceptable
quality. Details of the reduction procedure can be found in Udalski \etal
(1992).
\vskip4pt
In order to manipulate efficiently the huge amount of data the databases
for each observed field and color band were created. Photometry of each frame
with grade better than {\bf F} was aperture corrected and included in
the appropriate database. Details of database structure can be found in
Szyma{\'n}ski and Udalski (1993). Transformation to the standard {\it
VI} system is described in Udalski \etal (1992). The error of the zero
point of the photometry is not larger than 0.04~mag. The errors of
individual observations depend on magnitude of the star and frame grade
-- detailed analysis of errors can be found in Udalski \etal (1992) and
Udalski \etal (1994b). 
\vskip4pt
Over twenty ${15\arcm\times15\arcm}$ fields were observed in the OGLE
project. In this paper we present stars from the center of the Baade's
Window field designated as BWC. The J2000 coordinates of the BWC field
are: ${\alpha=18\uph03\upm24\ups}$, ${\delta=-30\arcd02\arcm00\arcs}$ 
(${l=1\zdot\arcd0}$, ${b=-3\zdot\arcd9}$).
\vskip4pt
First edition of the Catalog is limited to stars brighter than
${I=18.0}$~mag. In the following updates the Catalog will be extended to
fainter stars.  The Catalog has also an upper limit of brightness due to
saturation of bright star images in the CCD detector. This limit is
approximately ${14\upm}$ in $I$. The brightest star in the BWC Catalog
has ${I=13\zdot\upm85}$. The next part of the Catalog will also include a
table of brighter stars identified in the Baade's Window on a small number
of short exposure frames which were taken mainly for the purpose of
extending the upper parts of the color-magnitude diagrams (Udalski
\etal 1993a). It should be stressed, however, that these stars will not
constitute a homogeneous sample with the main part of the Catalog, due
to different procedure of reductions and small number of measurements.
\vskip4pt
The stars included in the catalog were extracted from the $I$-band
database. To enter the catalog, a star had to have at least 25 percent
of measurements of a "good quality". For the BWC field (total of 207
$I$ frames) this equals to 52 points. The measurement was defined as
"good" if it fulfilled the following criteria: the grade of the frame
was {\bf A} -- {\bf D}; the type of object assigned by the DoPHOT
program was "star-like"; there was no other object detected closer
than 0.75 pixel; the star was not too close to the frame edge;  the
magnitude error of this particular measurement was smaller than 3.0
times the upper sigma limit for a typical non-variable star of the same
brightness. The limiting curves "maximum sigma for non-variable stars
\vs magnitude" were derived from the analysis of the  distribution of
standard deviations of all stars (Udalski \etal 1993a).  From about
200\,000 objects in the database (roughly 44\,000 with mean
${I<18\upm}$) a total of 33196 stars fulfilled the above criteria and
entered the catalog. Table~1 lists brightness statistics of the BWC
field stars in one-magnitude bins. Fig.~1 presents all the selected
stars plotted in right ascension -- declination coordinates. The most
pronounced "hole" to the left of the center of this plot is a big
globular cluster (NGC~6522), where the density of stars is much too big
to detect individual objects. Other "holes" can be easily seen in
positions of bright stars which got overexposed on the frames.
Bigger dots in this figure mark the positions of all 212
periodic variable stars discovered by OGLE in the BWC field (Udalski
\etal 1994d).

\MakeTable{|l|c|c|c|c|}{12.5cm}{Number of BWC catalog stars in 
one-magnitude bins}
{
\hline\trule
$I$-magnitude           & 14.5$^*$  & 15.5 & 16.5 & 17.5\\[1mm]
\hline\trule
Number of stars         & 2089  & 5313 & 6616 & 19178 \\[1mm]
\hline
\multicolumn{5}{l}{\trule\small $^*$ including 48 stars brighter than 14\upm}
}

To derive mean ${V-I}$ color of the star, its $V$-band mean
magnitude was extracted from the $V$ filter database. In 1838 cases
$V$ magnitudes were unavailable because the star could not be
identified in the $V$ database. It is worth noting
that the $I$ and $V$ databases were constructed independently. The
cross identification of stars is based on second-order transformation
of pixel coordinates between the databases. The allowed difference
of the transformed $I$ database position and the $V$ database position
is 1.4 pixels. This value maximizes the number of matches (a match
requires that there is only one $V$ database object within a given
distance limit).

Equatorial coordinates of the variable stars were calculated using
transformation "frame position -- equatorial coordinates" derived using
stars from the HST Guide Star Catalog (Lasker \etal 1988). 21 GSC stars
were identified in the BWC field and used to determine transformation.
The relative accuracy of derived coordinates is about 0\zdot\arcs2 
while the absolute accuracy is worse -- about 1\arcs.

\section{The Catalog}
\vskip6pt
The General Catalog of Stars in the Galactic bulge contains the following
data:

\begin{enumerate}
\parskip 0mm
\itemsep 1mm
\item Star designation.

Each star is described as OGLE {\it field number},
where {\it field} is the general name of the OGLE field (see Table~1
in Udalski \etal 1994a) and {\it number} is a unique consecutive number
of the star in the given field. Initially the stars are
sorted according to their mean $I$ magnitude. Thus, lower number means
brighter star. This convention may be broken when the catalog is
updated. 
\item Right Ascention (J2000).
\item Declination (J2000).
\item $I$ mean magnitude.
\item Standard deviation ($\sigma_I$) of good $I$ band measurements.
\item Median photometry error 
(err$_{\mbox{\footnotesize med}}$) of a single good measurement.

$I$, $\sigma_I$ and err$_{\mbox{\footnotesize med}}$ values were 
calculated from good
measurements after two highest and two lowest magnitude points were
excluded.
\item $V$--$I$ estimate, if available.
\item Number of good photometry points ($N_I$).
\item Remarks, including references to the OGLE catalog of Periodic Variables 
and microlensing candidates.
\end{enumerate}

Table~2 presents an excerpt from the Catalog, namely 12 stars from the
beginning and 12 from the end of the whole sample. As none of these
stars do have any remarks, the 9th column of the table was omitted here
to ease formatting.

\MakeTable{|l|c|c|c|c|c|c|r|}{12.5cm}{The BWC catalog data}
{
\hline\trule
Star & R.A.(J2000) & Dec.(J2000) & $I$ & $\sigma_I$ & 
err$_{\mbox{\footnotesize med}}$ &  $V-I$ & $N_I$ \\[1mm]
\hline\trule
BWC 1    &  18\uph03\upm39\zdot\ups07 & -29\arcd57\arcm36\zdot\arcs8 & 13.847  &0.019 & 0.019 &  1.78 & 123 \\
BWC 2    &  18\uph03\upm41\zdot\ups14 & -29\arcd55\arcm59\zdot\arcs9 & 13.853  &0.021 & 0.019 &  1.69 & 119 \\
BWC 3    &  18\uph03\upm10\zdot\ups55 & -30\arcd00\arcm24\zdot\arcs8 & 13.865  &0.026 & 0.027 &  2.17 & 133 \\
BWC 4    &  18\uph03\upm06\zdot\ups26 & -30\arcd08\arcm59\zdot\arcs9 & 13.881  &0.017 & 0.013 &  2.09 &  79 \\
BWC 5    &  18\uph03\upm41\zdot\ups81 & -29\arcd58\arcm12\zdot\arcs0 & 13.882  &0.017 & 0.018 &  1.98 & 106 \\
BWC 6    &  18\uph03\upm14\zdot\ups76 & -30\arcd08\arcm14\zdot\arcs3 & 13.892  &0.020 & 0.014 &  1.85 &  70 \\
BWC 7    &  18\uph03\upm18\zdot\ups97 & -30\arcd06\arcm13\zdot\arcs0 & 13.893  &0.043 & 0.032 &  1.78 &  66 \\
BWC 8    &  18\uph03\upm28\zdot\ups41 & -30\arcd04\arcm57\zdot\arcs5 & 13.893  &0.020 & 0.024 &  1.68 & 138 \\
BWC 9    &  18\uph03\upm39\zdot\ups93 & -29\arcd57\arcm28\zdot\arcs8 & 13.894  &0.017 & 0.014 &  1.56 &  97 \\
BWC 10   &  18\uph03\upm35\zdot\ups68 & -30\arcd04\arcm38\zdot\arcs8 & 13.901  &0.015 & 0.019 &  2.00 & 135 \\
BWC 11   &  18\uph03\upm38\zdot\ups48 & -30\arcd01\arcm40\zdot\arcs0 & 13.907  &0.031 & 0.022 &  2.37 & 113 \\
BWC 12   &  18\uph03\upm01\zdot\ups79 & -30\arcd05\arcm16\zdot\arcs6 & 13.921  &0.025 & 0.022 &  2.28 & 140 \\[1mm]
 ...     &          ...               &             ...              &  ...   &  ...  &  ...  &   ... & ... \\[1mm]
BWC 33185 & 18\uph03\upm48\zdot\ups48 & -29\arcd55\arcm59\zdot\arcs3 & 17.991 & 0.095 & 0.048 &  1.14 & 186 \\
BWC 33186 & 18\uph02\upm46\zdot\ups52 & -30\arcd00\arcm53\zdot\arcs2 & 17.992 & 0.081 & 0.078 &  0.88 & 155 \\
BWC 33187 & 18\uph03\upm04\zdot\ups14 & -30\arcd05\arcm13\zdot\arcs8 & 17.992 & 0.150 & 0.082 &  1.39 & 142 \\
BWC 33188 & 18\uph02\upm59\zdot\ups79 & -30\arcd00\arcm06\zdot\arcs6 & 17.992 & 0.049 & 0.045 &  1.34 & 198 \\
BWC 33189 & 18\uph03\upm24\zdot\ups73 & -30\arcd05\arcm45\zdot\arcs8 & 17.992 & 0.073 & 0.074 &  1.24 & 198 \\
BWC 33190 & 18\uph03\upm28\zdot\ups71 & -29\arcd56\arcm21\zdot\arcs8 & 17.992 & 0.268 & 0.062 &  1.25 & 195 \\
BWC 33191 & 18\uph03\upm14\zdot\ups49 & -29\arcd56\arcm29\zdot\arcs9 & 17.993 & 0.261 & 0.054 &  1.30 & 198 \\
BWC 33192 & 18\uph03\upm40\zdot\ups72 & -29\arcd56\arcm01\zdot\arcs6 & 17.994 & 0.176 & 0.096 &  1.13 & 138 \\
BWC 33193 & 18\uph03\upm25\zdot\ups20 & -29\arcd55\arcm11\zdot\arcs2 & 17.995 & 0.177 & 0.053 &  1.09 & 198 \\
BWC 33194 & 18\uph03\upm38\zdot\ups26 & -30\arcd07\arcm11\zdot\arcs6 & 17.995 & 0.037 & 0.048 &  0.87 & 168 \\
BWC 33195 & 18\uph03\upm25\zdot\ups37 & -30\arcd03\arcm44\zdot\arcs2 & 17.996 & 0.466 & 0.157 &  0.54 &  87 \\
BWC 33196 & 18\uph03\upm30\zdot\ups17 & -29\arcd56\arcm57\zdot\arcs7 & 17.999 & 0.362 & 0.067 &  0.57 & 196 \\[1mm]
\hline
}

\section{Completeness of the Catalog}
It is important to have some information concerning completeness of
the Catalog. The detailed statistical analysis 
will be published elsewhere after next parts of the Catalog are
released. In this paper we present only a crude  estimate.

The main factor affecting the completeness of the catalog is determined
by the fraction of all stars which entered the whole selection process.
OGLE reduction procedure ignores stars which have bad photometry due to
CCD defects, severe blending or are overexposed {\it etc.\ }on the so
called "template" frame of a given field (Udalski \etal 1992). Such
objects are marked and are not measured on other frames. Thus a fraction
of objects is {\it a priori} missed. These objects had no chance
to be included in the Catalog.

The other factor comes from the adopted selection criteria, mainly from
the minimum number of good measurements. About one fourth of the
database objects brighter than 18\upm had too few good photometry 
points. We have checked, however, that this factor is not very
important, by running the selection process with much smaller lower
limit of good points. The test performed with this number set to 5\% of
all frames (for BWC it was 10 points) revealed only 4\% more objects
selected. This means that the objects rejected by this
criterium are, in vast majority, not true stars but rather artefacts of
chip defects, bleeding columns {\it etc.}

It is relatively easy to determine the first factor of completeness by 
randomly adding artificial stars to the frame and checking how many of
them have been recovered using standard reduction procedure. Such tests
have already been performed for analysis of CMDs of the Galactic bulge 
(Udalski \etal 1993a). The recovery efficiency for stars brighter than
${I=18\upm}$ has been found to be relatively flat and equal  to
${\approx80\%}$ for the densest BWC field and ${\approx 87\%}$ for the
least dense field TP8.

Independent estimate of this factor will be obtained and presented in
the next part of the Catalog describing the OGLE fields which overlap
the BWC field. We will check and analyze how many stars in the
overlapping regions would pass all criteria when processed independently
in two adjacent fields. In a few corner regions this analysis can be
done even for four fields.

The overlap analysis will provide also a better estimate of the second
factor, as we will be able to check how many of the stars which had too
few good points in one field did pass this criterium in the other field.

\section{Summary}
We present in this paper the first part of the OGLE General Catalog of
Stars in the Galactic bulge -- 33196 objects brighter than ${I=18\upm}$
detected in the BWC field. Catalog of stars found in 20 other fields
observed during the OGLE microlensing search will be published in
similar form in the forthcoming papers. The Catalog will be updated
regularly to include fainter stars, as well as when more data are
available or additional fields are observed.

The Catalog is available to astronomical community in electronic form
over the  Internet network using anonymous ftp service from {\it
sirius.astrouw.edu.pl} host (148.81.8.1), directory {\tt
/ogle/general\_catalog}. See README file in this directory for
details. The size of the file containing BWC field stars is 2.5\,MB
(0.5\,MB in the compressed form). For the availability of the Catalog
(as well as the Periodic Variables Catalog) on the CD-ROM and for
information about OGLE mirror archive in USA, consult our WWW
page: \newline
{\it http://www.astrouw.edu.pl/\~{~}ftp/ogle/ogle.html}.

{\bf Acknowledgements.} It is a great pleasure to thank B.\ Paczy\'nski
for valuable suggestions. This project was supported with the Polish KBN
grants 2P03D02508 to M.\ Szyma\'nski, 2P03D02908 to A.\ Udalski and  the
NSF grants AST 9216494 to B.\ Paczy\'nski, AST~9216830 to G.W.\
Preston.

\vspace{5mm}\noindent
FIGURE CAPTIONS:

\vspace{2mm}\noindent
Figure 1. BWC field catalog stars plotted in R.A. -- Dec. chart.
Bigger dots represent periodic variables found in the field (Udalski
\etal 1994d).

\end{document}